\input harvmac
\let\includefigures=\iftrue
\let\useblackboard=\iftrue
\newfam\black


\def\Title#1#2{\nopagenumbers\abstractfont\hsize=\hstitle\rightline{#1}%
\vskip 1in\centerline{\titlefont #2}\abstractfont\vskip .5in\pageno=0}
\def\Date#1{\vfill\leftline{#1}\tenpoint\supereject\global\hsize=\hsbody%
\footline={\hss\tenrm\folio\hss}}
%

\def\draftmode{\message{ DRAFTMODE }\def\draftdate{{\rm preliminary draft:
\number\day/\number\month/\number\year\ \ \hourmin}}%
\headline={\hfil\draftdate}\writelabels\baselineskip=20pt plus 2pt minus 2pt
 {\count255=\time\divide\count255 by 60 \xdef\hourmin{\number\count255}
  \multiply\count255 by-60\advance\count255 by\time
  \xdef\hourmin{\hourmin:\ifnum\count255<10 0\fi\the\count255}}}


\includefigures
\message{If you do not have epsf.tex (to include figures),}
\message{change the option at the top of the tex file.}
\input epsf
\def\figin{\epsfcheck\figin}\def\figins{\epsfcheck\figins}
\def\epsfcheck{\ifx\epsfbox\UnDeFiNeD
\message{(NO epsf.tex, FIGURES WILL BE IGNORED)}
\gdef\figin##1{\vskip2in}\gdef\figins##1{\hskip.5in}
\else\message{(FIGURES WILL BE INCLUDED)}%
\gdef\figin##1{##1}\gdef\figinbs##1{##1}\fi}
\def\DefWarn#1{}
\def\figinsert{\goodbreak\midinsert}
\def\ifig#1#2#3{\DefWarn#1\xdef#1{fig.~\the\figno}
\writedef{#1\leftbracket fig.\noexpand~\the\figno}%
\figinsert\figin{\centerline{#3}}\medskip\centerline{\vbox{
\baselineskip12pt\advance\hsize by -1truein
\noindent\footnotefont{\bf Fig.~\the\figno:} #2}}
\endinsert\global\advance\figno by1}
\else
\def\ifig#1#2#3{\xdef#1{fig.~\the\figno}
\writedef{#1\leftbracket fig.\noexpand~\the\figno}%
\global\advance\figno by1} \fi

\input color

\def\id{{1 \kern-.28em {\rm l}}}

\def\K3{{\bf K3}}
\def\journal#1&#2(#3){\unskip, \sl #1\ \bf #2 \rm(19#3) }
\def\andjournal#1&#2(#3){\sl #1~\bf #2 \rm (19#3) }

\def\tilde{\widetilde}

\def\frac#1#2{{#1\over#2}}

\def\inbar{\,\vrule height1.5ex width.4pt depth0pt}
\def\IC{\relax\hbox{$\inbar\kern-.3em{\rm C}$}}
\def\IR{\relax{\rm I\kern-.18em R}}
\def\IZ{\relax{\rm I\kern-.18em Z}}

%
%

%
\catcode`\@=11
\def\slash#1{\mathord{\mathpalette\c@ncel{#1}}}
\overfullrule=0pt

\def\NN{{\cal N}}

\def\underrel#1\over#2{\mathrel{\mathop{\kern\z@#1}\limits_{#2}}}

\catcode`\@=12


%

\def\det{{\rm det}}

\def\det{{\rm det}}



\lref\qheA{
Richard E. Prange and Steven M. Girvin, eds., 
``The Quantum Hall   Effect" (2nd ed.)
(Springer-Verlag, 1990).
}

\lref\qheB{
Sankar Das Sarma and Aron Pinczuk, eds., ``Perspectives in Quantum Hall Effects" (John Wiley and Sons, 1997).
}

\lref\qheC{
Steven M. Girvin, 
``The Quantum Hall Effect: Novel Excitations and Broken Symmetries", 
[arXiv:cond-mat/9907002].
}

\lref\sondhi{S.L. Sondhi, S.M. Girvin, J.P. Carini, D. Shahar, ``Continuous Quantum Phase Transitions'', arXiv:cond-mat/9609279}

\lref\huck{B. Huckestein, ``Scaling Theory of the Integer Quantum Hall Effect", Rev. Mod. Phys., {\bf 67}, 357, (1995)}

\lref\qheD{
 H.P. Wei {\it et.al.}, Phys. Rev. Lett. 61 (1988) 1294;

 P.T.~Coleridge, Phys. Rev.  B60, 4493 (1999);

 P.T.~Coleridge,  Solid State Comm. 112, 241 (1999).
 }

\lref\qheDa{
J. Wakabayashi, J. M. Yamane, and S. Kawaji, J. Phys. Soc. Jpn. 58, 1903 (1989); 

J. Wakabayashi, J. M. Yamane, and S. Kawaji, J. Phys. Soc. Jpn. 61, 1691 (1992);

M. D’Iorio, V. M. Pudalov, and S. M. Semenchinsky, High Magnetic Field in Semiconductor III, Quantum Hall Effect, Transport and Optics (Springer, Berlin, 1992), p. 56;

S. Koch, R. J. Haug, K. von Klitzing, and K. Ploog, Experiments on scaling in Alx Ga1−xAs/GaAs heterostructures under quantum Hall conditions. Phys. Rev. B {\bf 43}, 6828 (1991).

N. Q. Balaban, U. Meirav, and I. Bar-Joseph, Phys. Rev. Lett. {\bf 81}, 4967 (1998).
}

\lref\qheE{
 Li W.  {\it et.al.}, Phys. Rev. Lett. {\bf 94} 206807 (2005);
 Saeed K. {\it et.al.},  Phys. Rev. B {\bf 84} 155324 (2011);
}

\lref\BergmanGM{
  O.~Bergman, N.~Jokela, G.~Lifschytz and M.~Lippert,
  ``Quantum Hall Effect in a Holographic Model,''
JHEP {\bf 1010}, 063 (2010).
[arXiv:1003.4965 [hep-th]].
}

\lref\KutasovFR{
  D.~Kutasov, J.~Lin and A.~Parnachev,
  ``Conformal Phase Transitions at Weak and Strong Coupling,''
Nucl.\ Phys.\ B {\bf 858}, 155 (2012).
[arXiv:1107.2324 [hep-th]].
}

\lref\KutasovUQ{
  D.~Kutasov, J.~Lin and A.~Parnachev,
  ``Holographic Walking from Tachyon DBI,''
Nucl.\ Phys.\ B {\bf 863}, 361 (2012).
[arXiv:1201.4123 [hep-th]].
}

\lref\GoykhmanAZ{
  M.~Goykhman and A.~Parnachev,
  ``S-parameter, Technimesons, and Phase Transitions in Holographic Tachyon DBI Models,''
Phys.\ Rev.\ D {\bf 87}, no. 2, 026007 (2013).
[arXiv:1211.0482 [hep-th]].
}

\lref\KeskiVakkuriEB{
  E.~Keski-Vakkuri and P.~Kraus,
  ``Quantum Hall Effect in AdS/CFT,''
JHEP {\bf 0809}, 130 (2008).
[arXiv:0805.4643 [hep-th]].
}

\lref\DavisNV{
  J.~L.~Davis, P.~Kraus and A.~Shah,
  ``Gravity Dual of a Quantum Hall Plateau Transition,''
JHEP {\bf 0811}, 020 (2008).
[arXiv:0809.1876 [hep-th]].
}

\lref\KristjansenNY{
  C.~Kristjansen and G.~W.~Semenoff,
  ``Giant D5 Brane Holographic Hall State,''
JHEP {\bf 1306}, 048 (2013).
[arXiv:1212.5609 [hep-th]].
}

\lref\KristjansenHMA{
  C.~Kristjansen, R.~Pourhasan and G.~W.~Semenoff,
  ``A Holographic Quantum Hall Ferromagnet,''
JHEP {\bf 1402}, 097 (2014).
[arXiv:1311.6999 [hep-th]].
}

\lref\ZhangEU{
  S.~C.~Zhang,
  ``The Chern-Simons-Landau-Ginzburg theory of the fractional quantum Hall effect,''
Int.\ J.\ Mod.\ Phys.\ B {\bf 6}, 25 (1992)..
}

\lref\KobayashiSB{
  S.~Kobayashi, D.~Mateos, S.~Matsuura, R.~C.~Myers and R.~M.~Thomson,
  ``Holographic phase transitions at finite baryon density,''
JHEP {\bf 0702}, 016 (2007).
[hep-th/0611099].
}

\lref\KarchPD{
  A.~Karch and A.~O'Bannon,
  ``Metallic AdS/CFT,''
JHEP {\bf 0709}, 024 (2007).
[arXiv:0705.3870 [hep-th]].
}

\lref\OBannonIN{
  A.~O'Bannon,
  ``Hall Conductivity of Flavor Fields from AdS/CFT,''
Phys.\ Rev.\ D {\bf 76}, 086007 (2007).
[arXiv:0708.1994 [hep-th]].
}

\lref\BreitenlohnerBM{
  P.~Breitenlohner and D.~Z.~Freedman,
  ``Positive Energy in anti-De Sitter Backgrounds and Gauged Extended Supergravity,''
Phys.\ Lett.\ B {\bf 115}, 197 (1982)..
}

\lref\KaplanKR{
  D.~B.~Kaplan, J.~W.~Lee, D.~T.~Son and M.~A.~Stephanov,
  ``Conformality Lost,''
Phys.\ Rev.\ D {\bf 80}, 125005 (2009).
[arXiv:0905.4752 [hep-th]].
}

\lref\KarchSH{
  A.~Karch and E.~Katz,
  ``Adding flavor to AdS / CFT,''
JHEP {\bf 0206}, 043 (2002).
[hep-th/0205236].
}


\lref\NiemiRQ{
  A.~J.~Niemi and G.~W.~Semenoff,
  ``Axial Anomaly Induced Fermion Fractionization and Effective Gauge Theory Actions in Odd Dimensional Space-Times,''
Phys.\ Rev.\ Lett.\  {\bf 51}, 2077 (1983)..
}

\lref\RedlichKN{
  A.~N.~Redlich,
  ``Gauge Noninvariance and Parity Violation of Three-Dimensional Fermions,''
Phys.\ Rev.\ Lett.\  {\bf 52}, 18 (1984).
}

\lref\SemenoffDQ{
  G.~W.~Semenoff,
  ``Condensed Matter Simulation of a Three-dimensional Anomaly,''
Phys.\ Rev.\ Lett.\  {\bf 53}, 2449 (1984).
}

\lref\FujitaKW{
  M.~Fujita, W.~Li, S.~Ryu and T.~Takayanagi,
  ``Fractional Quantum Hall Effect via Holography: Chern-Simons, Edge States, and Hierarchy,''
JHEP {\bf 0906}, 066 (2009).
[arXiv:0901.0924 [hep-th]].
}

\lref\AlanenCN{
  J.~Alanen, E.~Keski-Vakkuri, P.~Kraus and V.~Suur-Uski,
  ``AC Transport at Holographic Quantum Hall Transitions,''
JHEP {\bf 0911}, 014 (2009).
[arXiv:0905.4538 [hep-th]].
}

\lref\GoldsteinAW{
  K.~Goldstein, N.~Iizuka, S.~Kachru, S.~Prakash, S.~P.~Trivedi and A.~Westphal,
  ``Holography of Dyonic Dilaton Black Branes,''
JHEP {\bf 1010}, 027 (2010).
[arXiv:1007.2490 [hep-th]].
}

\lref\RyuFE{
  S.~Ryu and T.~Takayanagi,
  ``Topological Insulators and Superconductors from String Theory,''
Phys.\ Rev.\ D {\bf 82}, 086014 (2010).
[arXiv:1007.4234 [hep-th]].
}

\lref\BayntunNX{
  A.~Bayntun, C.~P.~Burgess, B.~P.~Dolan and S.~S.~Lee,
  ``AdS/QHE: Towards a Holographic Description of Quantum Hall Experiments,''
New J.\ Phys.\  {\bf 13}, 035012 (2011).
[arXiv:1008.1917 [hep-th]].
}

\lref\KarchMN{
  A.~Karch, J.~Maciejko and T.~Takayanagi,
  ``Holographic fractional topological insulators in 2+1 and 1+1 dimensions,''
Phys.\ Rev.\ D {\bf 82}, 126003 (2010).
[arXiv:1009.2991 [hep-th]].
}

\lref\FujitaPJ{
  M.~Fujita,
  ``M5-brane Defect and QHE in $AdS_4 x N(1,1)/N=3 SCFT$,''
Phys.\ Rev.\ D {\bf 83}, 105016 (2011).
[arXiv:1011.0154 [hep-th]].
}

\lref\PalSX{
  S.~S.~Pal,
  ``Model building in AdS/CMT: DC Conductivity and Hall angle,''
Phys.\ Rev.\ D {\bf 84}, 126009 (2011).
[arXiv:1011.3117 [hep-th]].
}

\lref\JokelaNU{
  N.~Jokela, G.~Lifschytz and M.~Lippert,
  ``Magneto-roton excitation in a holographic quantum Hall fluid,''
JHEP {\bf 1102}, 104 (2011).
[arXiv:1012.1230 [hep-th]].
}

\lref\JokelaEB{
  N.~Jokela, M.~Jarvinen and M.~Lippert,
  ``A holographic quantum Hall model at integer filling,''
JHEP {\bf 1105}, 101 (2011).
[arXiv:1101.3329 [hep-th]].
}

\lref\RyuVQ{
  S.~Ryu, T.~Takayanagi and T.~Ugajin,
  ``Holographic Conductivity in Disordered Systems,''
JHEP {\bf 1104}, 115 (2011).
[arXiv:1103.6068 [hep-th]].
}

\lref\SemenoffYA{
  G.~W.~Semenoff and F.~Zhou,
  ``Magnetic Catalysis and Quantum Hall Ferromagnetism in Weakly Coupled Graphene,''
JHEP {\bf 1107}, 037 (2011).
[arXiv:1104.4714 [hep-th]].
}

\lref\BergmanRF{
  O.~Bergman, N.~Jokela, G.~Lifschytz and M.~Lippert,
 ``Striped instability of a holographic Fermi-like liquid,''
JHEP {\bf 1110}, 034 (2011).
[arXiv:1106.3883 [hep-th]].
}

\lref\BelhajFD{
  A.~Belhaj,
  ``On Fractional Quantum Hall Solitons in ABJM-like Theory,''
Phys.\ Lett.\ B {\bf 705}, 539 (2011).
[arXiv:1107.2295 [hep-th]].
}

\lref\KutasovFR{
  D.~Kutasov, J.~Lin and A.~Parnachev,
  ``Conformal Phase Transitions at Weak and Strong Coupling,''
Nucl.\ Phys.\ B {\bf 858}, 155 (2012).
[arXiv:1107.2324 [hep-th]].
}

\lref\JokelaSW{
  N.~Jokela, M.~Jarvinen and M.~Lippert,
  ``Fluctuations of a holographic quantum Hall fluid,''
JHEP {\bf 1201}, 072 (2012).
[arXiv:1107.3836 [hep-th]].
}

\lref\DavisAM{
  J.~L.~Davis and N.~Kim,
  ``Flavor-symmetry Breaking with Charged Probes,''
JHEP {\bf 1206}, 064 (2012).
[arXiv:1109.4952 [hep-th]].
}

\lref\BayntunBM{
  A.~Bayntun and C.~P.~Burgess,
  ``Finite Size Scaling in Quantum Hallography,''
[arXiv:1112.3698 [hep-th]].
}

\lref\FujitaFP{
  M.~Fujita, M.~Kaminski and A.~Karch,
  ``SL(2,Z) Action on AdS/BCFT and Hall Conductivities,''
JHEP {\bf 1207}, 150 (2012).
[arXiv:1204.0012 [hep-th]].
}

\lref\JokelaVN{
  N.~Jokela, G.~Lifschytz and M.~Lippert,
  ``Magnetic effects in a holographic Fermi-like liquid,''
JHEP {\bf 1205}, 105 (2012).
[arXiv:1204.3914 [hep-th]].
}

\lref\GrignaniQZ{
  G.~Grignani, N.~Kim and G.~W.~Semenoff,
  ``D7-anti-D7 bilayer: holographic dynamical symmetry breaking,''
Phys.\ Lett.\ B {\bf 722}, 360 (2013).
[arXiv:1208.0867 [hep-th]].
}

\lref\OmidVY{
  H.~Omid and G.~W.~Semenoff,
  ``D3-D7 Holographic dual of a perturbed 3D CFT,''
Phys.\ Rev.\ D {\bf 88}, no. 2, 026006 (2013).
[arXiv:1208.5176 [hep-th]].
}

\lref\BlakeTP{
  M.~Blake, S.~Bolognesi, D.~Tong and K.~Wong,
  ``Holographic Dual of the Lowest Landau Level,''
JHEP {\bf 1212}, 039 (2012).
[arXiv:1208.5771 [hep-th]].
}

\lref\KristjansenTN{
  C.~Kristjansen, G.~W.~Semenoff and D.~Young,
  ``Chiral primary one-point functions in the D3-D7 defect conformal field theory,''
JHEP {\bf 1301}, 117 (2013), [JHEP {\bf 1301}, 117 (2013)].
[arXiv:1210.7015 [hep-th]].
}

\lref\JokelaDW{
  N.~Jokela, J.~Mas, A.~V.~Ramallo and D.~Zoakos,
  ``Thermodynamics of the brane in Chern-Simons matter theories with flavor,''
JHEP {\bf 1302}, 144 (2013).
[arXiv:1211.0630 [hep-th]].
}

\lref\JokelaSE{
  N.~Jokela, M.~Jarvinen and M.~Lippert,
  ``Fluctuations and instabilities of a holographic metal,''
JHEP {\bf 1302}, 007 (2013).
[arXiv:1211.1381 [hep-th]].
}

\lref\MelnikovTB{
  D.~Melnikov, E.~Orazi and P.~Sodano,
  ``On the AdS/BCFT Approach to Quantum Hall Systems,''
JHEP {\bf 1305}, 116 (2013).
[arXiv:1211.1416 [hep-th]].
}

\lref\JokelaHTA{
  N.~Jokela, G.~Lifschytz and M.~Lippert,
  ``Holographic anyonic superfluidity,''
JHEP {\bf 1310}, 014 (2013).
[arXiv:1307.6336 [hep-th]].
}

\lref\BeaJXA{
  Y.~Bea, E.~Conde, N.~Jokela and A.~V.~Ramallo,
  ``Unquenched massive flavors and flows in Chern-Simons matter theories,''
JHEP {\bf 1312}, 033 (2013).
[arXiv:1309.4453 [hep-th]].
}

\lref\SonXRA{
  D.~T.~Son and C.~Wu,
  ``Holographic Spontaneous Parity Breaking and Emergent Hall Viscosity and Angular Momentum,''
JHEP {\bf 1407}, 076 (2014).
[arXiv:1311.4882 [hep-th]].
}

\lref\FilevBNA{
  V.~G.~Filev, M.~Ihl and D.~Zoakos,
  ``Holographic Bilayer/Monolayer Phase Transitions,''
JHEP {\bf 1407}, 043 (2014).
[arXiv:1404.3159 [hep-th]].
}

\lref\JokelaWSA{
  N.~Jokela, G.~Lifschytz and M.~Lippert,
  ``Flowing holographic anyonic superfluid,''
JHEP {\bf 1410}, 21 (2014).
[arXiv:1407.3794 [hep-th]].
}

\lref\MaganDWA{
  J.~M.~Magán, D.~Melnikov and M.~R.~O.~Silva,
  ``Black Holes in AdS/BCFT and Fluid/Gravity Correspondence,''
JHEP {\bf 1411}, 069 (2014).
[arXiv:1408.2580 [hep-th]].
}

\lref\WuDHA{
  C.~Wu and S.~F.~Wu,
  ``Hořava-Lifshitz gravity and effective theory of the fractional quantum Hall effect,''
JHEP {\bf 1501}, 120 (2015).
[arXiv:1409.1178 [hep-th]].
}

\lref\LippertJMA{
  M.~Lippert, R.~Meyer and A.~Taliotis,
  ``A holographic model for the fractional quantum Hall effect,''
JHEP {\bf 1501}, 023 (2015).
[arXiv:1409.1369 [hep-th]].
}

\lref\BeaYDA{
  Y.~Bea, N.~Jokela, M.~Lippert, A.~V.~Ramallo and D.~Zoakos,
  ``Flux and Hall states in ABJM with dynamical flavors,''
JHEP {\bf 1503}, 009 (2015).
[arXiv:1411.3335 [hep-th]].
}

\lref\LindgrenLIA{
  J.~Lindgren, I.~Papadimitriou, A.~Taliotis and J.~Vanhoof,
  ``Holographic Hall conductivities from dyonic backgrounds,''
[arXiv:1505.04131 [hep-th]].
}


\hfill TCDMATH-15-14
 \Title{} {\vbox{\centerline{A Holographic Model of Quantum Hall Transition}
}}

\bigskip

\centerline{\it  Andrea Mezzalira $^{1,2}$ and Andrei Parnachev $^{1,2}$}
\bigskip
\smallskip
\centerline{${}^{1}$ Hamilton Mathematics Institute and School of Mathematics,}
\centerline{Trinity College, Dublin 2, Ireland} 
\smallskip
\centerline{${}^{2}$Institute Lorentz for Theoretical Physics, Leiden University} 
\centerline{P.O. Box 9506, Leiden 2300RA, The Netherlands}
\smallskip

\vglue .3cm

\bigskip

\let\includefigures=\iftrue
\bigskip
\noindent
We consider a phenomenological holographic model, inspired by the D3/D7 system with a
2+1 dimensional intersection, at finite chemical potential and magnetic field.
At large 't Hooft coupling the  system is unstable and needs regularization;
the UV cutoff can be decoupled  by considering a certain double scaling limit.
At finite
chemical potential the model exhibits a phase transition between  states with filling fractions plus and minus one--half
as the magnetic field is varied.
By varying the parameters of the model,  this phase transition can be made to happen at arbitrary values
of the magnetic field. 

\bigskip

\Date{December 2015}

\newsec{Introduction and summary}

%
In condensed matter physics, the quantum Hall effect (QHE) is a general feature of 2+1 dimensional, low--temperature electron systems subject to strong magnetic field $B$ \refs{\qheA\qheB-\qheC}.
At zero temperature, by varying the magnetic field $B$, the transverse conductivity $\sigma_{xy}$ experiences sudden jumps between quantized values (plateaux)
\eqn\sigmaInt{\eqalign{
\sigma_{xy} = \nu \frac{e^2}{h}
\ ,
}}
where $\nu$ is the filling fraction, defined as the ratio of the charge density to the magnetic field, and it can assume integer (IQHE) or fractional values (FQHE).
Although the IQHE is well explained by considering localization-delocalization processes for free electrons moving in a random potential, a complete understanding of the fractional case, which relies on the strong interaction between electrons, is still lacking.
Remarkably, in both cases experiments show the presence of scaling behaviour with respect to the temperature.
Indeed, when the temperature $T$ is increased, the profile of the transition between plateaux is smoothed out and it is described by a power law of the temperature 
\eqn\sigmaIntB{\eqalign{
\frac{\partial \sigma_{xy}}{\partial B} 
\propto
 T^{-\kappa}
\ ,
}}
while at the same critical value of magnetic field the longitudinal conductivity exhibits sharp spikes.
Moreover,  the width of the region in which the transition occurs (or, equivalently, in which the longitudinal resistivity is different from zero) scales with the temperature
\eqn\expscalingB{\eqalign{
\Delta B
\propto
T^{\kappa} 
\ .
}}
The exponent $\kappa$ has been experimentally measured for different materials and between different pairs of plateaux (both in the integer and fractional case).
Initially, the same value $\kappa \sim 0.42$ had been found \refs{\sondhi\huck-\qheD} and this was interpreted as a signal of universal behaviour.
However, further investigations suggested that the value of $\kappa$ may be in general dependent on the experimental apparatus and the plateau transition considered \refs{\qheDa,\qheE} even if, concerning the IQHE, recent papers conjectured that the presence or absence of universality is affected by the range of the disorder potentials in the sample \qheE.

Due to the presence of strong interactions, it is difficult to understand the physics underneath the plateau transitions in the QHE.
Therefore, it would be interesting to have a holographic model of this phenomenon and to investigate the finite temperature
behaviour.
%
%
In this paper we focus on the phase transition at zero temperature, leaving the non--zero temperature analysis to future work.
There is a wide literature concerning the QHE and its holographic description.
Refs. \refs{\KeskiVakkuriEB,\DavisNV}\ studied quantum Hall plateaux using holographic D-brane constructions where the fermions are represented by open strings living on the 2+1 dimensional intersection of D3 and D7 system.
This approach was pursued further by several authors in various D-brane contexts \refs{\BergmanGM\KristjansenNY-\KristjansenHMA}.
Another interesting approach is based on the observation that some experimental results can be explained by a discrete duality group relating the different quantum Hall states.
Refs. \refs{\GoldsteinAW,\BayntunNX}\ and, more recently  \LippertJMA, considered a holographic model encoding this feature based on Einstein-Maxwell axion-dilaton action.
In this description, the quantum Hall states are represented by dyonic black holes and it is possible to capture the quantization of the Hall plateaux.
Other work on holographic quantum Hall physics includes \refs{\FujitaKW\AlanenCN\GoldsteinAW\RyuFE\BayntunNX\KarchMN\FujitaPJ\PalSX\JokelaNU\JokelaEB\RyuVQ\SemenoffYA\BergmanRF\BelhajFD\JokelaSW\DavisAM\BayntunBM\FujitaFP\JokelaVN\GrignaniQZ\OmidVY\BlakeTP\KristjansenTN\JokelaDW\JokelaSE\MelnikovTB\JokelaHTA\BeaJXA\SonXRA\JokelaWSA\MaganDWA\WuDHA\LippertJMA\BeaYDA-\LindgrenLIA}.

Although these attempts succeeded in explaining some of the features of QHE such as the presence of constant conductivity plateaux, the description of phase transitions between different quantum Hall plateaux remains elusive.
In this paper we  consider a holographic model that  exhibits
such a transition.
We follow the approach of  \refs{\KutasovFR\KutasovUQ\GoykhmanAZ-\KaplanKR} where the physics of interacting three-dimensional
fermions was argued to be holographically related to the physics of a tachyon field in the bulk of AdS space.
The three-dimensional fermions coupled to  four-dimensional $\NN=4$ super Yang Mills are realized as a low energy theory of the  D3/D7 branes configuration in which a small number of D7 branes intersects a large number of D3-branes along $2+1$ dimensions.
%
%
The holographic description involves finding a profile of the D7 brane propagating in the $AdS_5\times S^5$ space; there is a (below Breitenlohner-Freedman bound) tachyon mode which appears because the system is non-supersymmetric and unstable.
To understand the physics of this system, \KutasovFR\ proposed introducing a cutoff in the radial direction of AdS.
The model can be rendered renormalizable by taking cutoff to infinity and the tachyon mass to the BF value, while the physical scale remains fixed.

We study the consequences of having a finite chemical potential and a  finite magnetic field in this system.
The equations of motion, together with the regularity of the brane profile and the gauge field, give rise to two energetically inequivalent solutions.
Therefore, the system undergoes a first order phase transition precisely at $b=0$.
By computing the conductivities via linear response we observe that the transition is between two different plateaux, characterized by filling fractions $\nu=\pm\frac12$.
%
To obtain the phase transition for a non--zero value of the magnetic field $b_c$ we can phenomenologically modify  the action
(this involves explicit breaking of parity in three space--time dimensions).
Again, the phase transition is of the first order and it occurs between the two solutions of the equations of motion.
%
%
%

%
%

The remainder of the paper is organized as follows.
In section~2 the D3/D7 model is reviewed.
After discussing the Dirac--Born--Infeld action in the absence of gauge field, we consider the addition of the
Chern--Simons term and  turn on both the magnetic field and the charge density.
Then, we analyse the scaling symmetry of the full action and derive the charge density from the holographic dictionary.
Eventually we solve the equations of motion for different values of the magnetic field.
In section~3 we 
show that the system undergoes a phase transition between states with filling fraction $\nu=\pm\frac{1}{2}$ at $b_c=0$.
We also show how to change the value of $b_c$ by modifying the action.
The computation of the conductivities shows that the two phases exhibit two different values of the transverse conductivity.
Finally, in section~4 we discuss the results and  comment on future prospects. 
%


\newsec{The D$3$/D$7$ Model}


\subsec{Introduction}

We consider a brane system consisting of $N$ D$3$-branes intersecting a single D$7$-brane along a 2+1 dimensional defect.
The branes are oriented as follows
\eqn\branesetup{
\matrix{
 & t & x & y  & z & \rho & x_5 & x_6 & x_7 & x_8 & x_9
 \cr
 D3 & \bullet & \bullet & \bullet & \bullet 
 \cr
 D7 & \bullet & \bullet & \bullet & & \bullet & \bullet & \bullet & \bullet & \bullet &
 \cr 
}}
(Similar model has been used to analyse $\NN=4$ SYM with gauge group $SU(N)$  coupled to an $\NN=2$ superfield in the fundamental representation of $SU(N)$  
 \KarchSH.)
The gauge field lives in the $3+1$--dimensional Minkowski space--time labelled by the coordinates $\{t,x,y,z\}$ while the fermions are located in the $2+1$--dimensional defect $z=0$.
%
%
At strong coupling, the model is described by the D$7$-brane propagating  in the background $AdS_5\times S^5$ geometry generated by the stack of D$3$-branes.
In the probe limit, the back--reaction of the D$7$-brane on the background geometry is negligible.
%
%

To introduce the features relevant to the present work, this section is devoted to a brief review of the D$3$/D$7$ system.
The metric for the $AdS_5\times S^5$ geometry can be written as follows
\eqn\madss{\eqalign{
d s_{10}^2 
=
L^2 r^2  \left( 
		- d t^2
		+ d \vec{x}^2
	\right)
+
\frac{L^2}{r^2}
\left( 
	d \rho^2
	+
	\rho^2 d \Omega^2_4 +d x_9^2
\right)
\ ,
}}
where $L$ is the $AdS$ radius, $\vec{x}$ represents the boundary space directions $\{x,y,z\}$, the four--sphere is parametrized by the coordinates $\{x_4,\cdots,x_8\}$ and the following identities hold
\eqn\coordsA{\eqalign{
\rho^2 = \sum_{i=4}^{8}x_i^2
\ ,\qquad\qquad\qquad
r^2 = \rho^2 +x_9^2
\ .
}}
In this background the D$7$ probe brane wraps a four-sphere inside the $S^5$
and stretches along the $t,x,y$ directions.
The rest of the D$7$ world--volume is specified by a single embedding function 
$x_9=f(\rho)$, giving the following induced metric
%
%
%
\eqn\meightA{\eqalign{
d s_8^2 =
L^2 r^2  \left( 
		- d t^2
		+ d x^2 + d y^2 
	\right)
+
\frac{L^2}{r^2}
\left( 
	\left[ 1 + f'(\rho)^2 \right] d \rho^2
	+
	\rho^2 d \Omega^2_4
\right)
\ .
}}
The Dirac--Born--Infeld (DBI) action, up  to an overall constant, is
\eqn\actionZero{\eqalign{
S 
\simeq
\int_0^\Lambda d \rho 
\frac{\rho^4}{\rho^2 + f(\rho)^2} 
\sqrt{ 1 + f'(\rho)^2 } 
\ .
}}
The action \actionZero\ becomes quadratic in the regime of small $f(\rho)$;
it is clear that $f(\rho)$ is tachyonic with the tachyon mass below the Breitenlohner-Freedman bound \BreitenlohnerBM.
This is the holographic manifestation of the fact the the field theory is unstable.
As reviewed in detail in  \KutasovFR\ this instability is related to the formation
of the bi-fermion condensate at finite value of the 't Hooft coupling.
At infinitely large coupling the theory develops a gap of the order of the UV cutoff.
This can also be seen in holography by introducing the  UV cutoff $\Lambda$ in \actionZero.

In principle, the theory can be rendered renormalizable by taking the tachyon mass to the
BF value.
However we will not be doing this here: we believe that the main features of the model
relevant to the description of quantum Hall physics are not affected by this procedure.
A convenient way to think about the cutoff $\Lambda$ in  \actionZero\ is to imagine
that the model at that scale is modified and the UV physics lifts the tachyon mass above the BF
bound. 
Again, the details of this physics should not affect the infrared observables that we are after.\foot{
In \BergmanGM\ the tachyon mass is lifted by introducing the fluxes through the 
constituent two-spheres.}

The profile of $f(\rho)$ is obtained by solving the equations of motion, supplemented with the following boundary conditions
\eqn\bdyCZero{\eqalign{
f'(0) = 0
\ , \qquad\qquad\qquad
f(\Lambda) = 0
\ ,
}}
where the first one reflects the regularity of the brane at $\rho=0$ and the second one sets the fermion bare mass to zero.
Note that the scalar assumes a finite value at $\rho=0$.
It is useful to mention that the equations of motion derived from the action \actionZero\ preserve the scaling symmetry
\eqn\scalingZero{\eqalign{
\rho\rightarrow\alpha\rho
\ ,\qquad\qquad\qquad
f\rightarrow \alpha f
\ ,
}}
and this allows to chose the value of $f(0)$ arbitrarily.

One can also use a different coordinate system, defined as
\eqn\mapZ{\eqalign{
\rho = r \cos\theta
\ ,\qquad\qquad\qquad
x_9 = r \sin\theta
\ .
}}
Note that with the new choice of coordinates the action is manifestly symmetric with respect to the isometry group of $AdS_4$ $SO(2,3)$.
In Fig.~1 the map between the two coordinate systems and the profile of the D$7$-brane are shown.
\ifig\loc{Profile of D$7$-brane in the different coordinates systems. The scale $x_9(0) = r_0$ can be arbitrarily set to unity thanks to the scaling symmetry \scalingZero.}
{\epsfxsize5in\epsfbox{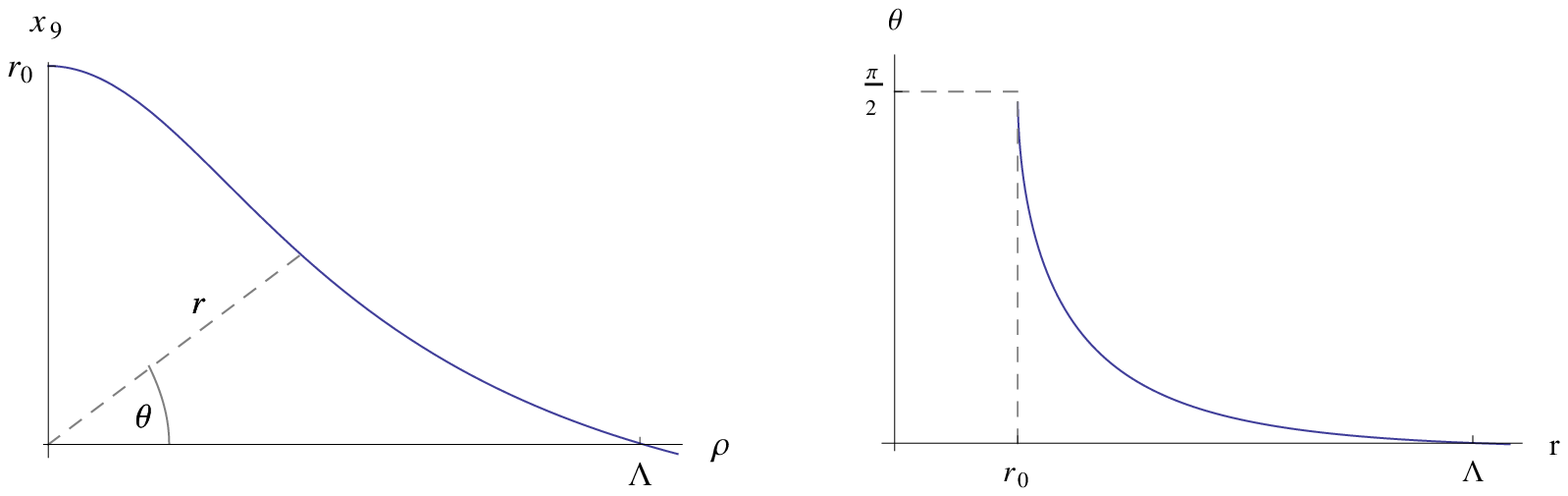}}
\noindent
The D-brane profile is now described by $\theta(r)$ and the action \actionZero\ becomes the tachyon DBI action
\eqn\actionZeroA{\eqalign{
S
\simeq
\int_{r_0}^\Lambda d r \,r^2 \, V(\theta(r)) \sqrt{1+ r^2 \theta'(r)^2} 
\ ,
}}
where the lower limit of the integral is defined as $r_0 = f(0)$.
%
%
The function $V(\theta(r))=\cos^4\theta(r)$ assumes the role of tachyon potential: the mass of the tachyon in the vacuum is obtained by expanding it up to second order in small $\theta$
\eqn\tacM{\eqalign{
V(\theta)\sim 1 + \frac{m^2}{2} \theta^2
\ ,
}}
that is, in our dimensionless units, $m^2 = -4 $. 
As stated previously, the value of the tachyon mass is below the Breitenlohner--Freedman bound $m_{BF}^2 = -d^2/4$, with $d$ number of boundary dimensions, therefore the scalar undergoes condensation.
The boundary conditions \bdyCZero\ are mapped to the following ones
\eqn\bdyCOne{\eqalign{
\partial_r \theta(r) |_{r_0} = - \infty
\ ,\qquad\qquad\qquad
\theta(\Lambda) = 0 
\ ,
}}
and $r_0=f(0)$ becomes
\eqn\bdyCTwo{\eqalign{
\theta(r_0)= \pm\frac{\pi}{2}
\ ,
}}
where the $\pm$ reflects the fact that the action is invarian under the transformation $\theta\rightarrow-\theta$ and therefore both the positive and negative profiles $\pm\theta(r)$ satify the equations of motion.
The potential $V(\theta(r))$ assumes monotonically the values from the maximum $V(\theta(\Lambda))=V(0)=1$ to the minimum $V(\theta(r_0))=V(\pm\pi/2)=0$.

The equation of motion for the action \actionZeroA\ reads
\eqn\EoMtZero{\eqalign{
&
\frac{
	r^2 \cos^3\theta(r)
}{
	\left( 
		1
		+
		r^2 \theta'(r)^2
	\right)^{3/2}
}
\left( 
	4 \sin\theta(r) 
	+
	4 r \cos\theta(r) \theta'(r)^2
	+
\right.
\cr
&
\left.\qquad
	+
	4 r^2 \cos\theta(r) \theta'(r)^3
	+
	r^2 \cos\theta(r) \theta''(r)
\right) = 0
\ .
}}
or, equivalently,
\eqn\EoMZeroZero{\eqalign{
\cos^4 \theta(r)\ \partial_r \left( 
	\frac{
		r^4 \theta'(r)
	}{
		\sqrt{
			1
			+
			r^2 \theta'(r)^2
		}
	}
\right)
=
\frac{
	r^2 \partial_\theta \left( 
		\cos^4 \theta(r)
	\right)
}{
	\sqrt{
		1
		+
		r^2 \theta'(r)^2
	}
}
\ .
}}
A term--wise study of the equation \EoMtZero\  near the point $r = r_0$ shows that the asymptotic expansion of the field $\theta(r)$ reads
\eqn\aSolZero{\eqalign{
\theta(r)
=
\pm
\left( 
\frac{\pi}{2} - \sqrt{\frac{10}{3}} \sqrt{r-r_0} + \ldots
 \right)
\ .
}}
In the following only the plus--solution will be considered and, thanks to the scaling symmetry \scalingZero, we choose $r_0=1$.
The result \aSolZero\ can also be obtained in a different coordinate system by solving the equations of motion for $x_9 = f(\rho)$ and then by using the change of coordinates \mapZ.
The near $\rho=0$ behaviour of the field $x_9(\rho)$ is described by
\eqn\rhoxNine{\eqalign{
x_9(\rho) = 1 - \frac{1}{5} \rho^2 + \ldots
\ ,
}}
which reduces to eq.~\aSolZero\ by means of the aforementioned map \mapZ.

To solve the equation \EoMZeroZero\ numerically it is convenient to consider a different set of boundary conditions
\eqn\bdyCThree{\eqalign{
\partial_r \theta(r) |_{r=1} = - \infty
\ ,\qquad\qquad\qquad
\theta(r=1)= \pm\frac{\pi}{2}
\ .
}}
The solutions in the two coordinate systems are shown in Fig.~1. 
The numerical value of UV cutoff $\Lambda$ is obtained by requiring $\theta(\Lambda)=0$ and it reads $\Lambda\sim 4.7305$, in perfect agreement with \KutasovFR.
%



\subsec{Adding magnetic field and charge density}

In this section we present the complete D$3$/D$7$ model studied in this work. 
The orientation of the branes is the same as in the setup \branesetup.
The background near--horizon metric for the D$3$--branes system reads
\eqn\dsone{
d s_{10}^2 
=
L^2 r^2  \left( 
		-  d t^2
		+ d \vec{x}^2
	\right)
+
\frac{ L^2 }{ r^2  } d r^2
+
L^2 d \Omega_5^2
\ ,
}
%
with $L$ the $AdS_5$ radius and the five--sphere is parametrized as
\eqn\Omegafive{\eqalign{
d \Omega_5^2 
&= d \theta^2 + \cos^2 \theta \, d \Omega_4^2
\cr
&=
d \theta^2 
+
\cos^2 \theta \, \left[ 
			d \phi_1^2 
			+
			\sin^2 \phi_1
			\left( 
				d \phi_2^2 
				+
				\sin^2 \phi_2
			\left( 
				d \phi_3^2 
				+
				\sin^2 \phi_3 d \phi_4^2				
			\right)
			\right)
		\right] 
		\ ,
}}
where $\{ \phi_1,\phi_2,\phi_3\}\in [0,\pi]$ and $\phi_4 \in [0,2\pi]$.
The coordinate $\theta$ can be defined in two different patches, covering each half a five--sphere:  $\theta \in [0,\pi/2]$ and $\theta \in [-\pi/2, 0,]$
.
In the followings we will consider the patch $\theta \in [0,\pi/2]$, while the other case will be commented later in this section.
The D$7$-brane extends along the $t\,,\,x\,,\,y$ and $r$ directions and wraps the $S^4$: its embedding is encoded in the two scalar fields $\theta(r)$ and $z(r)$.
For our purposes we set $z(r) = 0$, while, as already stated in the previous sections, $\theta(r)$ vanishes at the UV cutoff $r=\Lambda$ and assumes the value of $\pi/2$ at $r=r_0$. 
%

%
The background is supported by the following Ramond--Ramond (RR) five--form
\eqn\RRfive{\eqalign{
F_5
& = 
4 r^3 L^4 dt \wedge dx \wedge dy \wedge dz \wedge dr
	+
4 L^4 d\Omega_5	
\ ,
}} 
with
\eqn\dOmega{\eqalign{
d \Omega_5
& =
\cos^4 \theta \sin^3 \phi_1 \sin^2 \phi_2 \sin \phi_3 \,
d \theta \wedge d \phi_1 \wedge d \phi_2 \wedge d \phi_3 \wedge d \phi_4 
\cr
& \equiv 
\cos^4 \theta d \, \theta \wedge d \Omega_4
\ .
}} 
The RR four--form potential is defined as $d C_4 = F_5$ and in our conventions it reads
\eqn\RRfour{\eqalign{
C_4 
& =
r^4 L^4  d t \wedge d x \wedge d y \wedge d z
+
L^4 c(\theta) d \Omega_4
\ ,
}}
with 
\eqn\cdefA{\eqalign{
c(\theta)
=
\frac{3}{2} \theta
+
\sin\left( 2 \theta \right)
+
\frac{1}{8} \sin\left( 4 \theta \right)
+
c_1
\ .
}}
The constant of integration $c_1$ is fixed by the requirement that $c(\theta)d\Omega_4$ is a well defined differential form on the patch, namely its norm is not divergent in the whole domain considered.
The norm reads
\eqn\cnorm{\eqalign{
\left\|
c(\theta) d\Omega_4
\right\|^2
= 
\frac{c(\theta)^2}{\cos^8\theta}
\ .
}}
For $\theta\in[0,\pi/2]$, it is easy to see that in the neighbourhood of $\theta=\pi/2$ the norm blows up as
\eqn\cnormA{\eqalign{
\left\|
c(\theta) d\Omega_4
\right\|^2 
\sim 
\left( c_1 + \frac{3\pi}{4} \right)^2
\left( \frac{\pi}{2} - \theta  \right)^{-8}
+ \cdots
\ ,
}}
therefore, to cancel the divergence we fix the integration constant to be
\eqn\cone{\eqalign{
c_1 = - \frac{3\pi}{4} 
\qquad \Rightarrow \qquad
\left\|
c(\theta) d\Omega_4
\right\|^2 
=
\frac{16}{25}\left( \frac{\pi}{2} - \theta \right)^2
+\cdots
\ 
}}
and eq.~\cdefA\ becomes
\eqn\cdef{\eqalign{
c(\theta)
=
\frac{3}{2} \theta
+
\sin\left( 2 \theta \right)
+
\frac{1}{8} \sin\left( 4 \theta \right)
-
\frac{3\pi}{4}
\ .
}}
Thus, the norm of $c(\theta) d \Omega_4$ is defined everywhere for $\theta\in[0,\pi/2]$ and, in the same range of $\theta$, the function  $c(\theta)$ is always negative except at $\theta=\pi/2$, where it vanishes.\foot{
Our choice of function $c(\theta(r))$ \cdef\ differs from the one made in \BergmanGM\ by a gauge choice: there, the authors fix the constant of integration $c_1$ in eq.~\cdefA\ such that $c(\theta(r\rightarrow\Lambda))=c(0)=0$ and therefore $c(\theta(r_0))\neq 0$.
}
It is useful to remind that the field $\theta$ assumes the value $\pi/2$ when $r=r_0$
\eqn\ctheta{\eqalign{
c(\theta(r_0)) = c\left(\frac{\pi}{2}\right) = 0
\ .
}}
The case of $\theta \in [-\pi/2,0]$ is slightly different.
Indeed the request of a well defined norm for $c(\theta)d\Omega_4$ fixes differently the integration constant in \cdefA\ and we have
\eqn\cdefB{\eqalign{
c(\theta)_{\theta\in [-\pi/2,0]}
=
\frac{3}{2} \theta
+
\sin\left( 2 \theta \right)
+
\frac{1}{8} \sin\left( 4 \theta \right)
+
\frac{3\pi}{4}
\ .
}}
Therefore, the function $c(\theta)$ depends on the patch considered and note that the following relation holds
\eqn\cminusc{\eqalign{
c(-\theta)_{\theta\in [0,\pi/2]} = - c(\theta)_{\theta\in [-\pi/2,0]}
\ .
}}

As last ingredient, we introduce in the boundary theory a finite charge density and an external magnetic field $b$ (directed along $z$) by means of the gauge field, expressed in the Landau gauge
\eqn\Amu{\eqalign{
A 
& =
\frac{L^2}{2 \pi \alpha'}
\left( 
	a_0( r ) d t 
	+
	b \, x \, d y
\right)
\ .
}}

%
%

\subsec{Action}

Three terms contribute to the full action
\eqn\Stotzero{\eqalign{
S = S_{DBI} + S_{CS} + S_{bdy} \ ,
}}
that is, the Dirac--Born--Infeld action $S_{DBI}$, the Chern--Simons action $S_{CS}$ and the boundary term $S_{bdy}$.
The DBI action is defined as
\eqn\SDBIzero{\eqalign{
S_{DBI}
& = 
- T_7 \int d^8 \xi 
\sqrt{ - \det \left( G_{\alpha\beta} + 2\pi \alpha' F_{\alpha\beta} \right) }
\cr
& =
- 
\frac{8 \pi^2}{3} V_{1,2} T_7 L^8
\int dr \cos^4 \theta(r)
\sqrt{
\left( b^2 + r^4 \right)
\left[ 
	1 + r^2  \theta'(r)^2
	-
	a_0'(r)^2
\right]
}
\ ,
}}
where we have
\eqn\Vonetwo{\eqalign{
V_{1,2} \equiv \int dt \, dx \, dy
\ ,
\qquad
T_7=\left[(2\pi)^7 g_s (\alpha')^4  \right]^{-1}
\ ,
\quad
L^{4} = 4 \pi g_s N (\alpha')^2
\ ,
}}
with $g_s$ the string coupling, related with the ${\cal N}=4$ Yang--Mills coupling by $4 \pi g_s = g_{YM}^2$ and $N$ the number of D7--branes.
The volume for the four angular coordinates $\phi_i$ is
\eqn\Vfour{\eqalign{
V_4 
\equiv
\frac{8 \pi^2}{3} 
=
\left( 
	\prod_{i=1}^{3}
	\int_0^\pi d \phi_i
 \right)
 \int_0^{2\pi} d \phi_4 \sin^3 \phi_1 \sin^2 \phi_2 \sin \phi_3
 \ ,
}}
while the induced metric $G_{\alpha\beta}$ on the probe D$7$-brane is
\eqn\dseight{\eqalign{
ds_8^2 
& =
r^2 L^2 \left( 
	- dt^2 + dx^2 + dy^2
\right)
+
L^2 \left( 
	\frac{1}{r^2} + \theta'( r )^2
\right) dr^2
+
L^2 \cos^2\theta \,d\Omega_4^2
\ .
}}
The Chern-Simons contribution reads\foot{The sign convention is the same of \BergmanGM .}
\eqn\SCSzero{\eqalign{
S 
& =
\frac{\left( 2 \pi \alpha' \right)^2 T_7}{2} 
\int_{\cal{M}} F_5 \wedge F \wedge A
\ ,
}}
which, after integrating by parts, generates two terms.
We will refer to the first one as the Chern-Simons term
\eqn\SCS{\eqalign{
S_{CS} 
& =
- 
\frac{\left( 2 \pi \alpha' \right)^2 T_7}{2} 
\int_{\cal{M}} P[C_4] \wedge F \wedge F
\cr 
& =
\frac{8 \pi^2}{3} V_{1,2} T_7 L^5 
\int dr \, c(\theta(r)) b \, a_0'(r)
\ ,
}}
while the second is a boundary term
\eqn\Sbdy{\eqalign{
S_{bdy} 
& =
\frac{\left( 2 \pi \alpha' \right)^2 T_7}{2} 
\int_{\partial\cal{M}} P[C_4] \wedge F \wedge A
\cr 	
& =
- \frac{8 \pi^2}{3} V_{1,2} T_7 L^5 
c(\theta(\Lambda)) b \, a_0 (\Lambda)
\ ,
}}
where the relevant part of pullback of the RR four--form potential is obtained from eq.~\RRfour
\eqn\CPfour{\eqalign{
P[C_4] = L^4 c(\theta(r)) d \Omega_4 
\equiv
L^4 c(\theta(r)) d \Omega_4 
\ ,
}}
with $c(\theta)$ given in \cdef\ or \cdefB, depending on which patch is considered.
Note that the Chern--Simons action \SCSzero\ is invariant under the gauge transformation of the RR potential four--form $C_4$
\eqn\cgauge{\eqalign{
C_4 \rightarrow C_4 + \alpha_4
\ ,
}}
where $\alpha_4$ is a closed form $d \alpha_4 = 0$.
%
%
The full gauge invariant action can be written as
\eqn\fullS{\eqalign{
S =
&
{\cal N}
\int dr 
\left[  
- V( \theta(r) )
\sqrt{
\left( b^2 + r^4 \right)
\left[ 
	1 + r^2  \theta'(r)^2
	-
	a_0'(r)^2
\right]
}
+
c(\theta(r)) b a_0'(r)
\right]
\cr
&
-
{\cal N} c(\theta(\Lambda)) b \, a_0 (\Lambda)
\ ,
}}
with
\eqn\calN{\eqalign{
{\cal N} = \frac{8 \pi^2}{3} V_{1,2} T_7 L^8 \ ,
}}
and where, as in eq.~\actionZeroA,\ we introduce the tachyonic potential $V(\theta(r))=\cos^4(\theta(r))$.
%


\subsec{Equations of Motion}

In this section we solve the equations of motion obtained from the action \fullS\ for the scalar field $\theta(r)$ and the gauge field $a_0(r)$.
Since the action depends on $a_0(r)$ only through its derivative $a_0'(r)$, the equation of motion for the gauge field can be written as
\eqn\EoMaZero{\eqalign{
b c (\theta(r)) 
+
\frac{
	V(\theta(r)) \sqrt{
		b^2 + r^4
	}
}{
	\sqrt{
		1 
		+
		r^2  \theta'(r)^2
		-
		a_0'(r)^2
	}	
} a_0'(r)
=
d
\ ,
}}
with $d$ integration constant. 
From this equation we can derive the consistency condition for the flux of electric field on the D$7$-brane.
Indeed, since the D$7$-brane does not touch the Poincar\'e horizon at $r=0$, the flux has to vanish.
Said otherwise, there are no sources of electric field.
Following \KobayashiSB, we clarify this statement by requiring the norm of the electric displacement field to be non--singular
%
%
%
\eqn\electricD{\eqalign{
\left\|
	\frac{1}{\sqrt{-g}} \frac{\delta S}{\delta F_{rt}} 
\right\|^2 
\propto
\frac{d^2}{\cos^8 \theta}
\ ,
}}
which leads to the condition
\eqn\dandbA{\eqalign{
d = 0 \ .
}}
%
%
The presence of the Chern-Simons term \SCS, as it will be explained in the following section, gives rise to a non--zero charge density.
Therefore, we can consider a solution with a non--zero charge density.
%
%

We can solve algebraically eq.~\EoMaZero, obtaining
\eqn\EoMa{\eqalign{
a_0'(r)
= 
-
\frac{
	\sqrt{ 1 + r^2  \theta'(r)^2 } 
}{
	\sqrt{ b^2 c(\theta(r))^2 + \left( b^2 + r^4 \right) V(\theta(r))^2  }
}
b c(\theta(r))
\ .
}}
In the same way we compute the equation of motion for $\theta(r)$.
We have
\eqn\EoMpsiV{\eqalign{
&
\partial_r
\left( 
	\frac{
		r^2  V(\theta(r))
		\sqrt{
			b^2 + r^4
		}
	}{
		\sqrt{
			1 
			+
			 r^2 \theta'(r)^2
			-
			a_0'(r)^2 
		}		
	}\theta'(r)
\right)
+
\cr 
& 
-
\left( 
	\partial_{\theta(r)} V(\theta(r))
\right)
\sqrt{ b^2 + r^4}
\sqrt{
	1 + r^2  \theta'(r)^2
	-
	a_0'(r)^2 
} 
+
\cr
& 
+
\left( 
	\partial_{\theta(r)}c(\theta(r))
\right)
b \, a_0'(r) = 0
\ .
}}
The equations of motion \EoMpsiV\  and \EoMa\ show the following scaling symmetry
\eqn\scalingA{\eqalign{
	r \rightarrow \alpha r
\ ,\qquad
	\theta \rightarrow \theta
\ ,\qquad
	b \rightarrow \alpha^2 b
\ ,\qquad
	a_0' \rightarrow a_0'
\ ,
}}
and from eq.~\fullS\ it is easy to see that the action scales as follows
\eqn\Sscaling{\eqalign{
S \rightarrow \alpha^3 S 
\ .
}}
%
%
%
%
%

Finally, we solve numerically the equations of motion \EoMaZero\  and \EoMpsiV\ in the zero temperature regime but with non--zero magnetic field.
To do so, we need a more precise set of boundary conditions than \bdyCThree.
An accurate analysis of the equations of motion near the point $r = r_0$. shows that the $\theta(r)$ field behaves as
\eqn\asyNZero{\eqalign{
\theta(r) 
\simeq
\pm 
\left(
	\frac{\pi}{2} 
	-
	\sqrt{\frac{10}{r_0}}
	\sqrt{\frac{b^2 + r_0^4}{b^2 + 3 r_0^4}}
	\sqrt{r - r_0}
\right)
\ .
}}
Performing a scan over different values of $b$ we obtain the results in Fig.~2 and Fig.~3.
In the figures we considered only the solutions obtained considering the condition $\theta(r)\sim \pi/2$ near $r_0$ of eq.~\asyNZero, however, as we have already stated, the equations of motion are invariant under $\theta\rightarrow-\theta$ and therefore the solution starting from $\theta(r_0)=-\pi/2$ can be found just by changing the sign of the profile of $\theta(r)$ we presented here.
\ifig\loc{(a) Tachyon potential $V(\theta)$; (b) $c(\theta)$ function; (c) Three solutions of the equation of motion for various values of the magnetic field: $b=0$, $0.5$ and $1$. The red curve represents $b=0$. 
All solutions are ultimately rescaled to satisfy $\theta(\Lambda)=0$.
}
{\epsfxsize5.7in\epsfbox{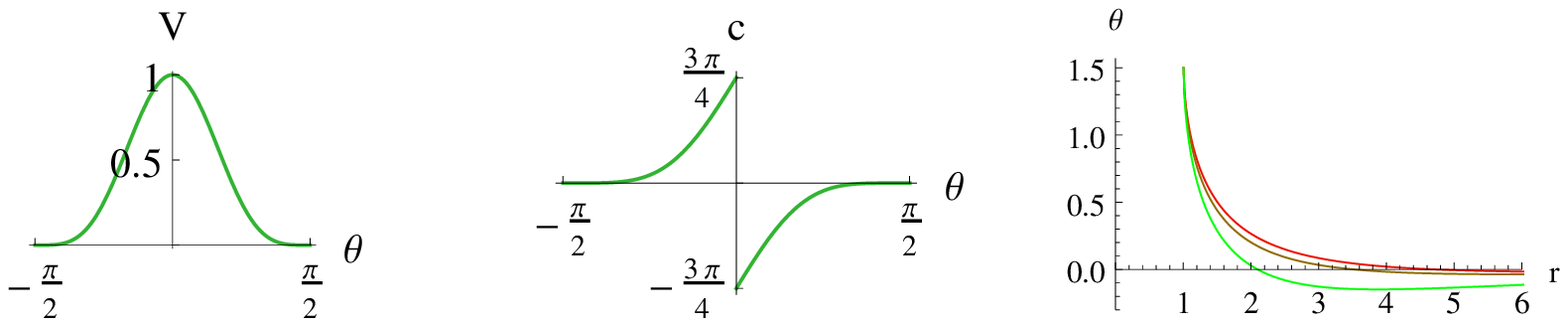}}
Increasing $|b|$, the solutions intersect the $\theta=0$ axis for values of the radial coordinates closer to $r=r_0=1$.
However their behaviour remains qualitatively the same, as shown in details in Fig.~3. 
\ifig\loc{Details of the three solutions of Fig.~2 (from red to green: $b=0$, $0.5$ and $1$).  All solutions are ultimately rescaled to satisfy $\theta(\Lambda)=0$.
}
{\epsfxsize5.3in\epsfbox{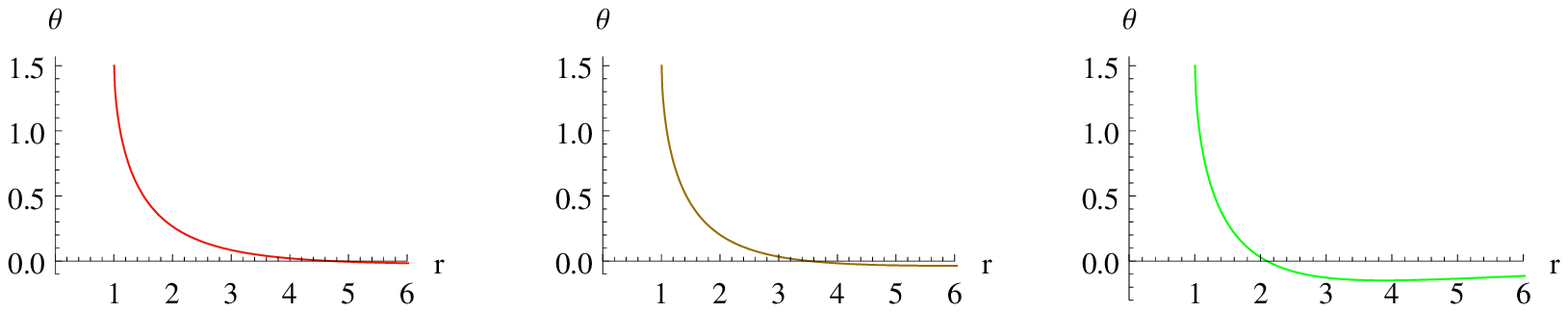}}

\subsec{Computing the charge density}

To find the favored state we have to compare the Gibbs free energies computed on the different solutions.
We keep the chemical potential fixed and therefore we study the system in the grand--canonical ensemble by considering the Gibbs free energy.
By the holographic dictionary, the Gibbs free energy is associated with the on--shell (euclidean) action $S$
\eqn\Gibbs{\eqalign{
\Omega (\mu, T, b) 
& = 
- 
\left( 
	S_{DBI} + S_{CS}
\right)_{on-shell}
+
{\cal N} c(\theta(\Lambda))\,b\,a_0(\Lambda)
\ ,
}}
The chemical potential $\mu$ is defined, as usual, as the value of the temporal component of the gauge field \Amu\ at the cutoff $\Lambda$
\eqn\muB{\eqalign{
\mu = A_t(\Lambda) = \frac{L^2}{2 \pi \alpha'}  a_0(\Lambda)
\ ,
}}
while the charge density $\rho$ is its conjugate variable
\eqn\chargeD{\eqalign{
\rho 
=
-\frac{1}{V_{1,2}} \frac{\delta \Omega}{\delta \mu} 
\ .
}}
To compute $\rho$ we perform a variation at constant $b$ of the Gibbs free energy \Gibbs
\eqn\ttA{\eqalign{ 
\delta \Omega  = 
-\int_{r_0}^\Lambda d r 
\frac{
	\delta ({\cal L}_{DBI} + {\cal L}_{CS})
}{
	\delta \partial_r a_0
}
\delta \left( \partial_r a_0 \right) 
-
\frac{
	\delta S_{bdy}
}{
	\delta a_0
} \delta a_0 (\Lambda)
\ .
}}
The first term evaluates as
\eqn\dB{\eqalign{
\frac{
	\delta ({\cal L}_{DBI} + {\cal L}_{CS})
}{
	\delta \partial_r a_0
} 
= d \ ,
}}
and it vanishes as in eq.~\dandbA.
B noting that eq.~\muB\ implies $\delta\mu = \frac{L^2}{2\pi\alpha'}\delta a_0(\Lambda)$ and by observing that $\delta a_0(0) = 0$, eq.~\ttA\ reduces to
\eqn\ttB{\eqalign{
\delta \Omega 
= 
\frac{\delta S^E_{bdy}}{\delta a_0} \delta a_0 (\Lambda)
= 
\frac{2\pi\alpha'}{L^2}\frac{\delta S^E_{bdy}}{\delta a_0} \delta \mu 
\ ,
}}
namely, the complete action $S$ depends on $a_0(\Lambda)$ only through the boundary term.
Comparing eq.~\ttB\ with the first law of thermodynamics $\delta \Omega= - V_{1,2}\rho \, \delta \mu$ allows us to write
\eqn\chargeD{\eqalign{
\rho 
=
-
\frac{2\pi\alpha'}{L^2} \frac{{\cal N}}{V_{1,2}} b\, c(\theta(\Lambda)) 
= 
- 
\left( \frac{2\pi\alpha'}{L^2} \right)^2 \frac{{\cal N}}{V_{1,2}} B\, c(\theta(\Lambda))
\ ,
}}
where from the definition of the gauge field eq.~\Amu\ we have the physical external magnetic field $B=\frac{L^2}{2\pi\alpha'} b$.
By using the definitions of ${\cal N}$ eq.~\calN, and of the $AdS$ radius and D--brane tension eq.~\Vonetwo\ we have
\eqn\chargeDa{\eqalign{
\rho 
=
-\frac{4}{3} \frac{ N }{ (2\pi)^2} B c(\theta(\Lambda))
\ ,
}}
where $N$ is the number of D7--brane (which has been set to $1$).
From the charge density we can define the filling fraction $\nu$ as 
\eqn\fillingf{\eqalign{
\nu 
=
-
\frac{2\pi}{N} \frac{\rho}{B}
=
-
\frac{2}{3\pi} c(\theta(\Lambda))
=
\pm \frac{1}{2}
\ ,
}}
where the plus sign corresponds to the patch $\theta\in[0,\pi/2]$ and the minus to $\theta\in[-\pi/2,0]$.
%
%
Note that under the scaling transformation of eqs.~\scalingA\ and \Sscaling\  the charge density behaves as the magnetic field $b$
\eqn\scalingrho{\eqalign{
\rho \rightarrow \alpha^2 \rho
\ ,
}}
and therefore the filling fraction $\nu$ is invariant under the rescaling \scalingA.
Note that the filling fraction (plus or minus) one-half is consistent with the parity anomaly
of a three-dimensional fermion coupled to the gauge field, as already noticed in \DavisNV.


\subsec{Rescaling and Normalizing the Free Energy}

The computation of the Gibbs free energy requires the introduction of a UV cutoff $\Lambda$.
%
One can always use the scaling symmetry to make any solution $\theta(r)$ satisfy $\theta(\Lambda)=0$.
%
%
It is important to note that in general the value of $r$ in which $\theta$ vanishes changes by considering different values of the magnetic field $b$.
Since we want to compare the different Gibbs free energies we need to have the same cutoff for every given value of $b$ and therefore we rescale the physical quantities accordingly.
The cutoff $\Lambda$ is chosen as the value of $r$ for which $\theta(r)=0$ in the case of zero magnetic field $b=0$.
The rescaling is made possible thanks to the scaling symmetry \scalingA\ and \Sscaling\  with the parameter set as
\eqn\Sparam{\eqalign{
\alpha = \frac{\Lambda}{r_*(b)} \ ,
}}
where $r_*(b)$ is defined by $\theta(r_*(b))=0$.
Of course, also the chemical potential has to be held fixed for all the different values of magnetic field.
The boundary value of $a_0(r_*(b))$ is obtained by means of the scale symmetry \scalingA  
\eqn\muLambda{\eqalign{
a_0 (r_*(b)) \frac{\Lambda}{r_*(b)} = \mu_0
\ ,
}}
with $\mu_0$ the external, fixed value for the chemical potential.
Finally, since we are only interested in the difference between the Gibbs free energies, we choose to normalize $\Omega$ by subtracting the value $\Omega_0$, namely the Gibbs free energy at zero $b$.
Therefore, the magnetic field and the Gibbs free energy after the rescaling and the normalization read
\eqn\scalingB{\eqalign{
b \rightarrow \left(  
	\frac{\Lambda}{r_*(b)} 
\right)^2 b
\ ,\qquad\qquad
\Omega \rightarrow \left(  
	\frac{\Lambda}{r_*(b)} 
\right)^3 \Omega - \Omega_0
\ .
}}

\subsec{Black--Hole embedding profiles}

The solution we have considered so far is characterized by the fact that it stops at a certain value of the radial coordinate $r=r_0$ and, therefore, it does not enter the Poincar\'e horizon, which is defined in our coordinate system as the line $r=0$ in the $\{r,\theta\}$ plane.
However, in general we could expect that the equations of motion can be satisfied also by profiles of $\theta(r)$ which go up to $r=0$.
We will refer to this  kind of solution as black hole (BH) embedding (since the D$7$--brane crosses the horizon in $r = 0$), while the old solution, stopping at $r=r_0$, corresponds to a Minkowski (MN) embedding (the D$7$--brane ends smoothly outside the horizon).
Since BH embeddings enter the Poincar\'e horizon, they allow the presence of an electric flux and therefore the requirement \electricD\ no longer holds.
In this section we study the presence of such profiles and we find two new solutions: $\theta=0$ and another non--trivial $\theta(r)$.
By analysing their Gibbs free energies we conclude that for small values of the chemical potential these new profiles are less favored than the MN solution.

At first we consider the constant solution $\theta=0$.
The equation of motion for the gauge field \EoMaZero\ reads
\eqn\EoMaZeroCase{\eqalign{
a_0'(r)
= 
\frac{
	( d - b c(0) )
}{
	\sqrt{ (d-b c(0))^2 + b^2 + r^4 }
}
=
\frac{
	\tilde{d}
}{
	\sqrt{ {\tilde{d}}^2 + b^2 + r^4  }
}
\ ,
}}
where we defined the constant $\tilde{d}=d - b c(0)$, while the one for the $\theta(r)$ \EoMpsiV\ becomes
\eqn\EoMpsiZeroCase{\eqalign{
b\, a_0'(r) = 0 \ .
}}
When $b=0$ eq.~\EoMpsiZeroCase\ is satisfied automatically.
By integrating eq.~\EoMaZeroCase\ and from the requirement that the gauge field $a_0(r)$ vanishes at $r=0$ we have
\eqn\muZeroCase{\eqalign{
\mu = \int_0^\Lambda 
\frac{
	d
}{
	\sqrt{ d^2 + r^4 }
}
\ ,
}}
which allows us to fix the constant $d$ as a function of the chemical potential $\mu=a_0(\Lambda)$.
We note that a maximum value of $\mu$ exists: as $\mu$ approaches this value (determined by the UV-cutoff) $d$ goes to infinity.
However, when the magnetic field is different from zero eqs.~\EoMaZero\ and \EoMpsiZeroCase\ imply $\tilde{d}=0$.
In other words, $\theta=0$ is not a solution when both $b$ and $\mu$ are non--vanishing.

The other possible (BH) solution has a non--trivial profile $\theta(r)$.
%
It is found by analysing numerically the equations of motion ({\it c.f.} Fig.~4)
\ifig\loc{Another (BH) solution of the equations of motion for $b=0$, $d=.4$ and $b=1$, $d=.1$.
}
{\epsfxsize4.3in\epsfbox{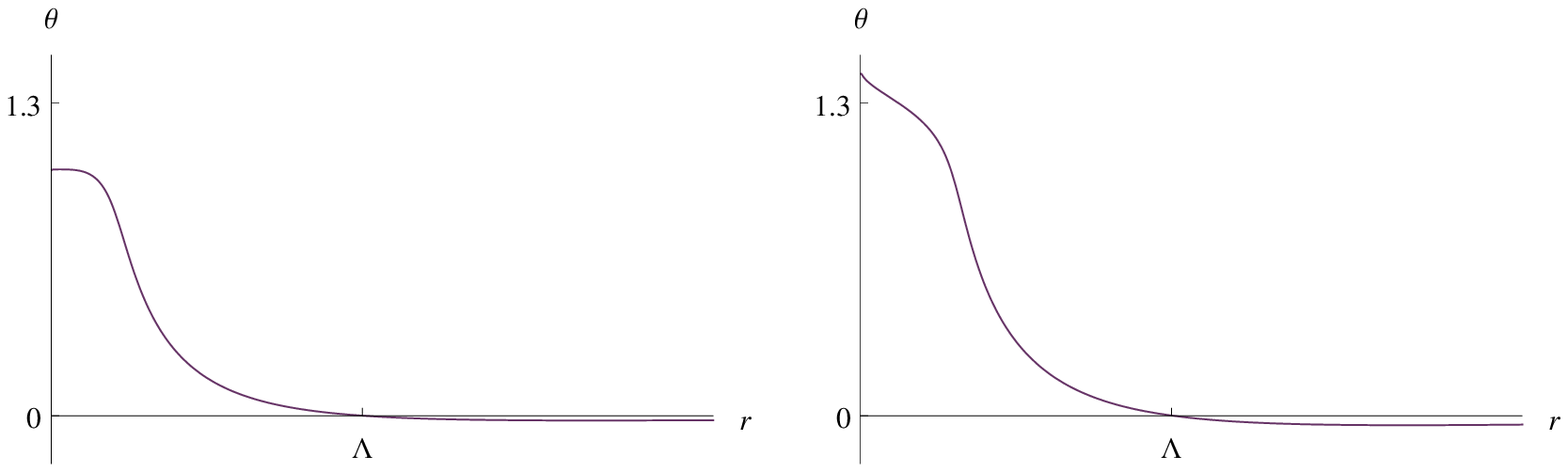}}
Note that $\theta(r)$ vanishes at the UV-cutoff $\Lambda$.
As in the $\theta(r)=0$, $b=0$ case, the constant $d$ of eq.~\EoMaZero\ is obtained by integrating the equation of motion \EoMaZero\ and it is fixed by the chemical potential.

At zero magnetic field, by comparing the Gibbs free energies $\Omega$ as functions of $\mu$ we find that for small values of the chemical potential the favored solution is the MN one. 
%
%
For $\mu$ larger than a critical value $\mu_c$ (but below the maximal value of $\mu$ discussed above), the BH solution becomes the favored one, while the $\theta=0$ solution never has the lowest energy.
The phase diagram slightly changes when $b$ is increased: for small $\mu$ the MN profile is always favored and, as before, after a critical value of $\mu$ the interpolating solution dominates.
The difference with respect to the $b=0$ case is that, as we stated previously, the constant $\theta=0$ profile does not exist anymore.
These results are represented in Fig.~5.
\ifig\loc{Plot of the Gibbs free energy as a function of the chemical potential for the three different kinds of solution, for $b=0$ (left panel) and $b=1$ (right panel).
The solid blue line represents the MN profile, the orange short--dashed one the BH embedding and the purple, large--dashed curve is associated with the $\theta=0$ solution. 
Note that $\theta=0$ is not a solution of the equations of motion if $b\neq0$.
}
{\epsfxsize4.3in\epsfbox{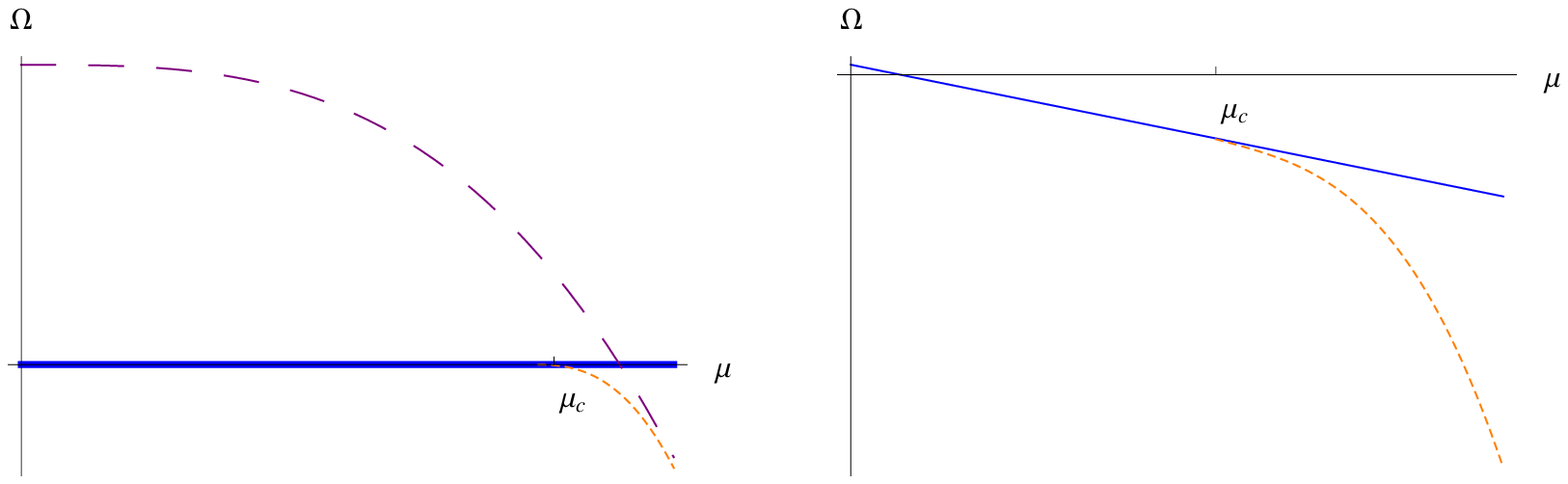}}
To summarize, we found that the the model we are considering admits profiles which enter the Poincar\'e horizon $r=0$.
However, in a certain range of $\mu$ (below a critical value $\mu_c$) these new solutions have larger Gibbs free energy than the MN profile.
Therefore we can limit ourselves to study only the MN solution.

\subsec{Conductivities}

To complete the analysis of the holographic model of the quantum Hall effect  we compute the longitudinal and transverse component of the conductivity $\sigma$, defined as
\eqn\sigmaDef{\eqalign{
J_i = \sigma_{ij} E_j \ ,
}}
where $J_i$ is the electric current induced in the medium and $E_i$ is the external electric field.
To introduce in the model these new features, we modify the gauge field \Amu\ as follows 
\eqn\AmuFluc{\eqalign{
A =
\frac{L^2}{2 \pi \alpha' } 
	\left[
		a_0(r) dt +
		\left( 
			e\, t + a_x(r)
		\right) dx
		+
		\left( 
			b\, x + a_y(r)
		\right) dy
\right]
\ ,
}}
namely we add a background electric field and the fluctuations along the longitudinal and transverse directions $a_x(r)$ and $a_y(r)$ respectively,
As in the case of magnetic field explained in section~2.5, the physical electric field is defined as
\eqn\Efield{\eqalign{
E_x = 
\frac{L^2}{2 \pi \alpha' } 
e
\ ,
\qquad\qquad\qquad
E_y=0
\ 
}}
The action \fullS\ is then modified as follows
\eqn\fullSVcon{\eqalign{
S =
&
{\cal N}
\int dr 
\left[  
- V(\theta(r))
\sqrt{
	Y
}
+
c(\theta(r)) \left( 
	b \, a_0'(r) + e \, a_y'(r)	
\right)
\right]
\cr
&
-
{\cal N} c(\theta(\Lambda)) \left( 
	b \, a_0 (\Lambda) + e \, a_y (\Lambda)
\right)
\ ,
}}
where we defined
\eqn\conY{\eqalign{
Y =
&
\left( 
	b^2 -e^2 + r^4
\right)\left( 
	1 + r^2 \theta'(r)^2
\right)
-
\left( 
	b^2 + r^4
\right) a_0'(r)
\cr
&
+
r^4 a_x'(r)^2
+
\left( 
	r^4 -e^2
\right) a_y'(r)
-
2 b\,e\,a_0'(r) a_y'(r)
\ .
}}
In analogy with section~$2.3$ and in particular with eq.~\EoMaZero\ the action depends only on the derivatives of the gauge field, therefore the equations of motion for the fluctuations $a_0$, $a_x$ and $a_y$ read, respectively
\eqn\EoMaCond{\eqalign{
&
\frac{ V(\theta(r)) }{ \sqrt{Y}} 
\left( 
	\left( r^2 + b^4 \right) a_0'(r)
	+ 
	b\, e\, a_y'(r)
\right)
+
b\, c( \theta(r) ) = d
\ ,
\cr
&
\frac{ V(\theta(r)) }{ \sqrt{Y}} 
r^4 a_x'(r) = d_x
\ ,
\cr
&
\frac{ V(\theta(r)) }{ \sqrt{Y}} 
\left( 
	- 
	\left( r^2 + b^4 \right) a_y'(r)
	+ 
	b\, e\, a_0'(r)
\right)
+
e\, c( \theta(r) ) = d_y
\ .
}}
With the same argument of regularity of the electric displacement field \electricD\ of sec.~2.3, the constants of integration $d$, $d_x$ and $d_y$ are set to zero. 
The AdS/CFT dictionary prescribes that the physical currents $J_x$ and $J_y$ are computed by differentiating the action \fullSVcon\ with respect to the boundary values of the gauge field $A_x(\Lambda)= \frac{L^2}{2 \pi \alpha' } a_x(\Lambda) $ and $A_y(\Lambda)=\frac{L^2}{2 \pi \alpha' } a_y(\Lambda)$.
As in the computation of the charge density, the only contribution to the action \fullSVcon\ of the boundary values of the gauge field is through the boundary term.
%
%

Therefore, in both patches, the results for the currents read
\eqn\currents{\eqalign{
J_x = 0
\ ,
\qquad\qquad
J_y = 
-
\frac{4}{3} \frac{N}{ (2 \pi)^2} E_x\, c(\theta(\Lambda)) \ ,
}}
From eq.~\sigmaDef\ and the expression of the filling fraction $\nu$ eq.~\fillingf, we obtain the longitudinal and Hall (transverse) conductivities
\eqn\condFin{\eqalign{
\sigma_{xx} = 0
\ ,
\qquad\qquad\qquad
\sigma_{xy} = -\sigma_{yx} = \frac{N \nu}{2 \pi}
\ .
}}
%
%
The longitudinal conductivity vanishes while, since $c(0)$ is constant, we see the emergence of the plateaux for the transverse conductivity.
Note that, since we are in a zero--temperature regime, the transition is not smoothed out.
%


\newsec{Tachyon Model and Phase Transitions}

In this section we study the Gibbs free energies of the two sets of solutions we found previously, namely the one starting from $\theta= \pi/2$ and the one starting from $\theta=-\pi/2$.
In particular, we analyse the competition between the two solutions to understand whether a phase transition occurs by varying the magnetic field, namely if the difference between the Gibbs free energies changes sign.

\subsec{Transition at $b=0$}

By considering the equation of motion \EoMa\ it is easy to see that both the DBI \SDBIzero\ and the Chern--Simons \SCS\ terms are invariant under $\theta\rightarrow-\theta$.
However, since the chemical potential $\mu=a_0(\Lambda)$ has to be kept fixed, this symmetry ceases to hold when we consider the boundary term \Sbdy.
Therefore the whole action \fullS\ and the Gibbs free energy $\Omega$ \Gibbs\ are not invariant and there is a competition between the solution starting from $\theta=\pi/2$ and the one starting from $\theta=-\pi/2$.
This results in a phase transition located exactly at $b_c=0$.
At zero magnetic field the boundary term $S_{bdy}$ vanishes and the two solutions have the same energy.
For a non--zero $b$ the boundary term computed on the two solutions, being linear in $b\, c(0)$, assumes opposite values, and there is a phase transition at $b=0$.
%
%

To understand the phase transition in a more quantitative way we proceed as described in section~2.5: after computing the on--shell euclidean action \Gibbs\ (that is, the Gibbs free energy $\Omega$) we rescale both $b$ and $\Omega$ as in \scalingB\ in order to have the same cutoff $r=\Lambda$ for every $b$ considered.
Then, we renormalize the Gibbs free energy by subtracting the value of $\Omega$ at $b=0$.
By following the computation of the conductivity in section~2.8 we realize that, since the phase transition is between solutions defined in different patches, the transverse conductivity $\sigma_{xy}\propto c(0)$ \condFin\ assumes different values.
Therefore, at $b=b_c=0$ we find a plateau transition.
The results are shown in Fig.~6.
\ifig\loc{Results from the D--brane construction. Left: the difference between the Gibbs free energy for the two solutions; right: plot of the transverse conductivity \condFin\ as a function of the magnetic field.
}
{\epsfxsize5.3in\epsfbox{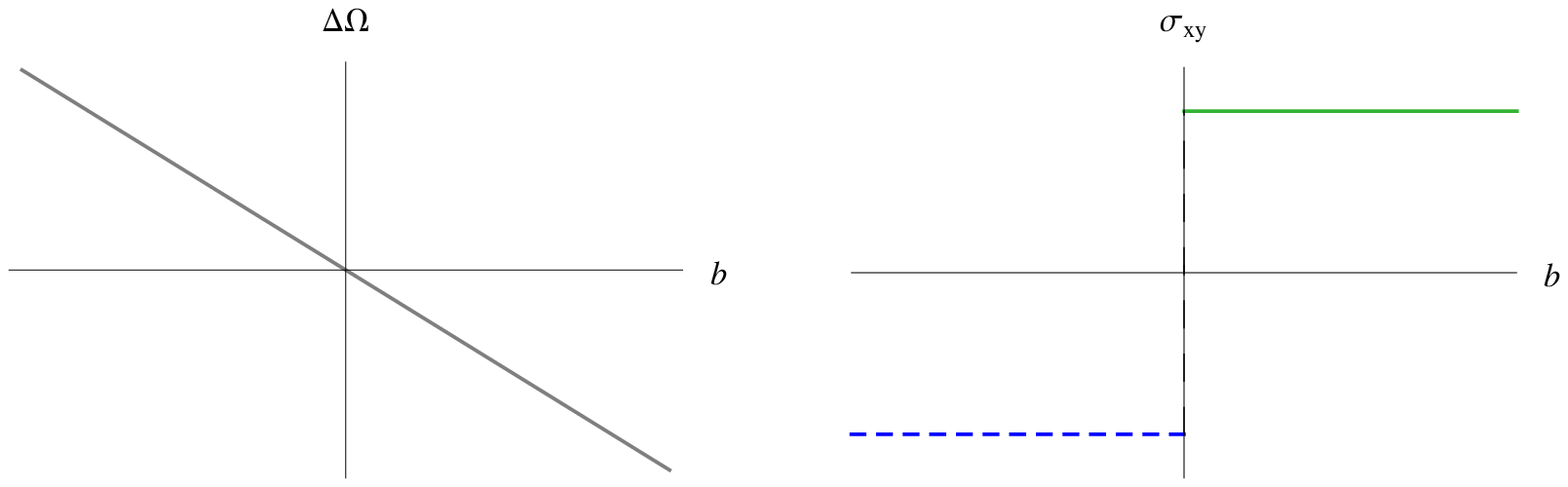}}

\subsec{Model for transitions at $b\neq0$}

As reminded in the introduction, the QHE occurs in presence of a strong magnetic field.
Therefore, the phase transition at $b_c=0$ is not satisfactory from the point of view of describing a real--world quantum Hall transition.
In order to model a phase transition at a non--zero value of the magnetic field we adopt a bottom--up approach by modifying the action \fullS.
More in detail, we modify the tachyon potential $V(\theta)$ and the $c(\theta)$ function in the domain $\theta \in [-\pi/2,0]$.
%
For simplicity, we ask that the new potential preserves the extreme points of the original potential $V(\theta)=\cos^4 \theta$, namely the unstable vacuum $\theta=0$ and the stable miminum $\theta=\pi/2$. 
Moreover, we want to keep fixed the mass of the tachyon ({\it c.f.} eq.~\tacM) and therefore we ask the expansion of the potential near $\theta=0$ ({\it c.f.} eq.~\tacM) to be the same, up to second order
\eqn\Vexp{\eqalign{
V(\theta) 
\sim
1 - 2 \theta^{2}\ .
}}
To fulfill the previous requirements and, at the same time, to generate a new phase we 
%
modify the behaviour of the potential for $\theta\in[-\pi/2,0]$.

In section.~2.2 we showed that both the tachyon potential and the $c(\theta)$ function are derived from the geometry of the five sphere $S^5$.
More precisely, the tachyon potential is related to the metric \Omegafive\ while the RR four--form is determined by the volume element $d\Omega_5$ \dOmega.
These two quantities can be expressed in terms of the single function, for instance the tachyon potential.
Indeed we have
\eqn\OmegafiveDefo{\eqalign{
d \Omega_5^2
=
d \theta^2
+
\sqrt{V(\theta)} d \Omega_4^2
\ ,
}}
and
\eqn\dOmegaDefo{\eqalign{
d \Omega_5
=
V(\theta) d\theta \wedge d\Omega_4
\ .
}}
The $c(\theta)$ function is then obtained by integrating the RR five--form \RRfive, with the $S^5$ volume form given by \dOmegaDefo, in analogy with the derivation of section~2.2.
Note that the regularity condition \ctheta, namely $c\left(\pi / 2\right) = 0 = c\left(-\pi/2\right)$ still holds.
Thanks to these considerations, the modification of the two functions $V(\theta)$ and $c(\theta)$ can be loosely interpreted as a deformation of the bulk geometry in the domain $\theta\in[-\pi/2,0]$.
Of course an honest string construction of this type requires more work.

We consider the following $V(\theta)$
\eqn\Vnew{\eqalign{
V(\theta)=\cos^4 \theta \left( 1 - 3 \sin^4(2\theta) e^{-4\theta^2} \right)
\ ,
}}
while $c(\theta)$ is given by
\eqn\cnewtwo{\eqalign{
c(\theta) 
=
\int_{-\pi/2}^\theta 4 V(\theta') d \theta' 
\ .
}}
The profiles are shown in Fig.~7. 
We note that with the choice made the value of $c(\theta)$ in $\theta=0$ is different from the original one \cdefB.
\ifig\loc{left and center: profiles for the tachyon potential \Vnew\ $V(\theta)$ and its associated RR four--form $c(\theta)$ \cnewtwo.
%
%
Right: profile for the alternative RR four--form (3.7), not derived from the potential \Vnew.
}
{\epsfxsize5.3in\epsfbox{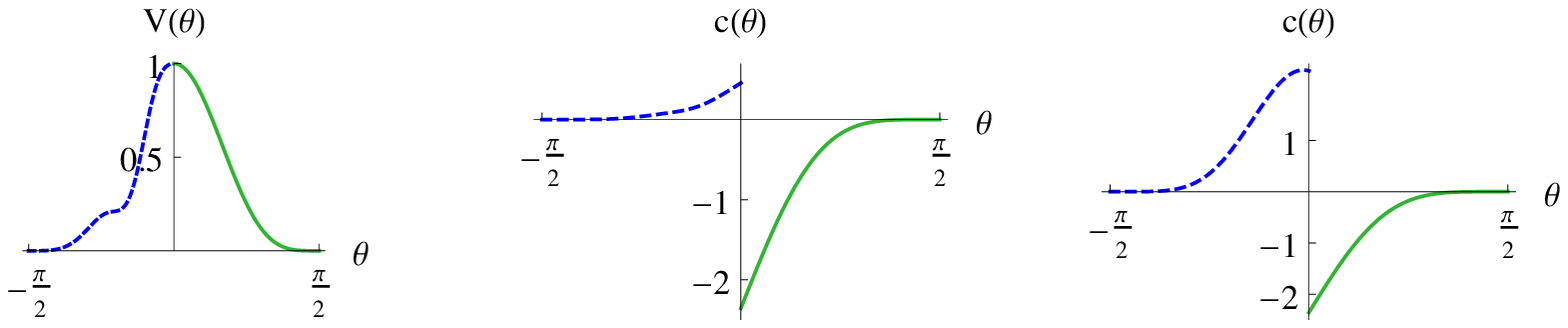}}
We then solve numerically the equations of motion in the two patches $\theta\in[-\pi/2,0]$ and $\theta\in[0,\pi/2]$ for different values of $b$.
In Fig.~8 we draw the corresponding results.
\ifig\loc{Details of three solutions. From left (red) to right (green): $b=0$, $0.5$ and $1$. All solutions are ultimately rescaled to satisfy $\theta(\Lambda)=0$.
}
{\epsfxsize5.3in\epsfbox{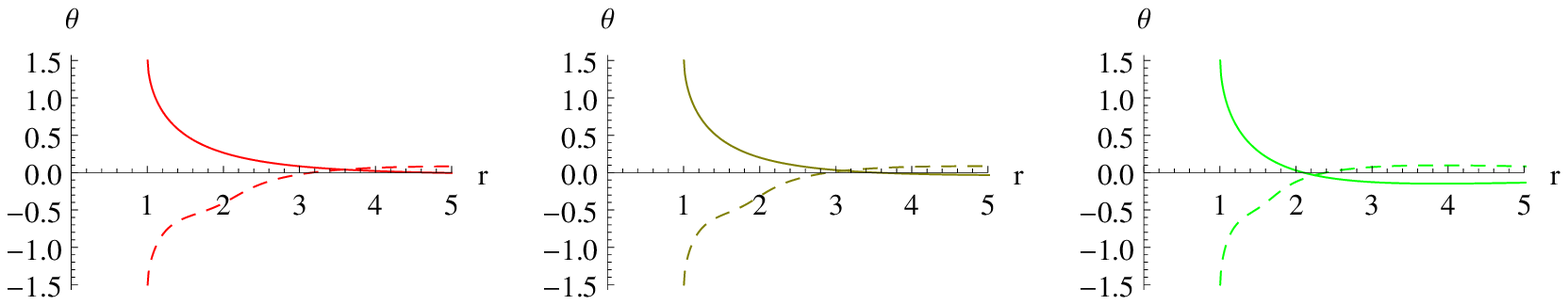}}
Finally, we compute the Gibbs free energy $\Omega$, as described in the previous sections.
We observe that the three contributions to the Gibbs free energy behave differently with increasing $b$ ({\it c.f.} Fig.~9, central panel): the DBI terms are almost constant while the Chern-Simons increases while the boundary term, as we have already pointed out in the previous section, is linear in $b$.
Since at $b=0$ only the DBI term is different from zero, the particular shape of the potential modifies the initial value of the Gibbs free energy.
Then, at $b>0$, the combined effect of both $V(\theta)$ and $c(\theta)$ changes the slope of $\Omega$, triggering a first order phase transition at $b_c>0$: below the critical value of $b$ we see that the new potential is favored, while above the solution associated with the original potential is preferred.
\ifig\loc{Results for eq.~\Vnew\ as tachyon potential and $c(\theta)$ obtained from $V(\theta)$. Left: the difference between the Gibbs free energy for the two solutions; center: the various contributions to the difference of Gibbs free energy: the DBI (small dashed line), the Chern-Simons (solid line) and the boundary term (large dashed line) for the two solutions; right: plot of the transverse conductivity \condFin\ as a function of the magnetic field.
}
{\epsfxsize5.3in\epsfbox{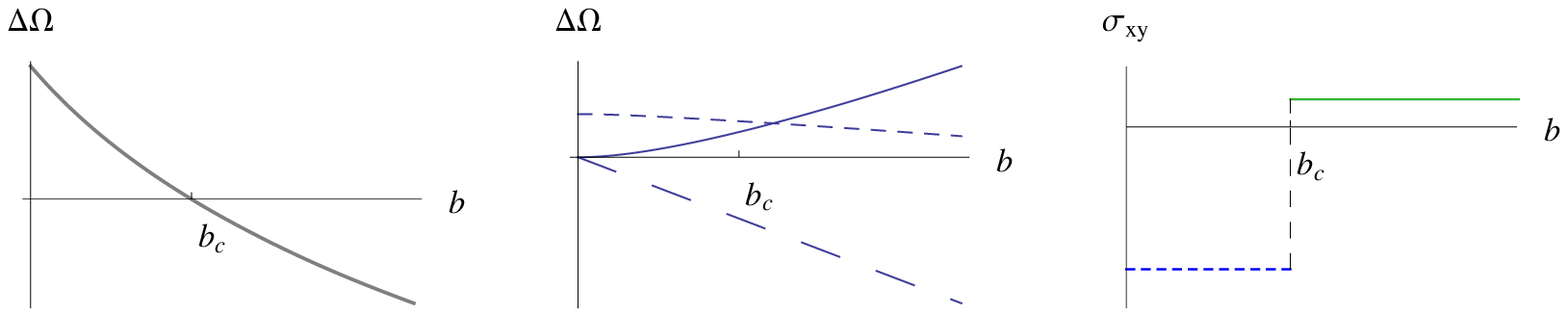}}
The computation of section~2.8 shows that for $b=b_c$ the transverse conductivity \condFin\ jumps between two different plateaux (right panel of Fig.~9) characterized by the values of the $c(\theta)$ in the two different patches at $\theta=0$.
%

We can investigate if it is possible to modify the action further to obtain other values for the transverse conductivity.
For instance, we can demand $\sigma_{xy}$ to jump between two opposite values, as in the case of section.~3.1.
To do so in the $\theta\in[-\pi/2,0]$ patch we demand
\eqn\cdefN{\eqalign{
c(0) = + \frac{3\pi}{4}
\ .
}}
To fulfill this condition we have to treat the tachyon potential and RR four--form as independent functions.
We choose to keep the $V(\theta)$ of eq.~\Vnew\ and we define a new $c(\theta)$ as follows
\eqn\cnew{\eqalign{
c(\theta)
& =
\left(
	\frac{3}{2} \theta 
	+ 
	\sin(2\theta)
	+
	\frac{1}{8} \sin(4\theta)
	+
	\frac{3\pi}{4} 
\right)
\left( 
	1
	-
	\sin(2 \theta)
\right)
\ ,
}}
and its profile is shown in the right panel of Fig.~7.
Again, by solving the equations of motion (the profiles are shown in Fig.~10) and by computing the Gibbs free energies we obtain a first order phase transition for $b=b_c\neq0$.
As expected, at $b=b_c$ the system experiences a transition between two plateaux in the transverse conductivity $\sigma_{xy}$ characterized by opposite values.
The results are shown in Fig.~11.
\ifig\loc{Details of three solutions. From left (red) to right (green): $b=0$, $0.5$ and $1$. All solutions are ultimately rescaled to satisfy $\theta(\Lambda)=0$.
}
{\epsfxsize5.3in\epsfbox{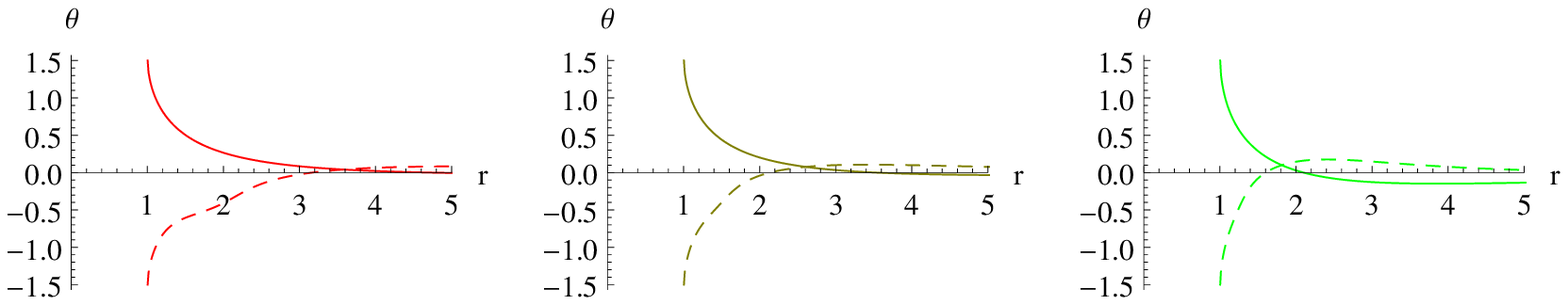}}
\ifig\loc{Results for the model with tachyon potential given by \Vnew\ and $c(\theta)$ given by \cnew. Left: the difference between the Gibbs free energy for the two solutions; center: the various contributions to the difference of Gibbs free energy: the DBI (small dashed line), the Chern-Simons (solid line) and the boundary term (large dashed line) for the two solutions; right: plot of the transverse conductivity \condFin\ as a function of the magnetic field.
}
{\epsfxsize5.3in\epsfbox{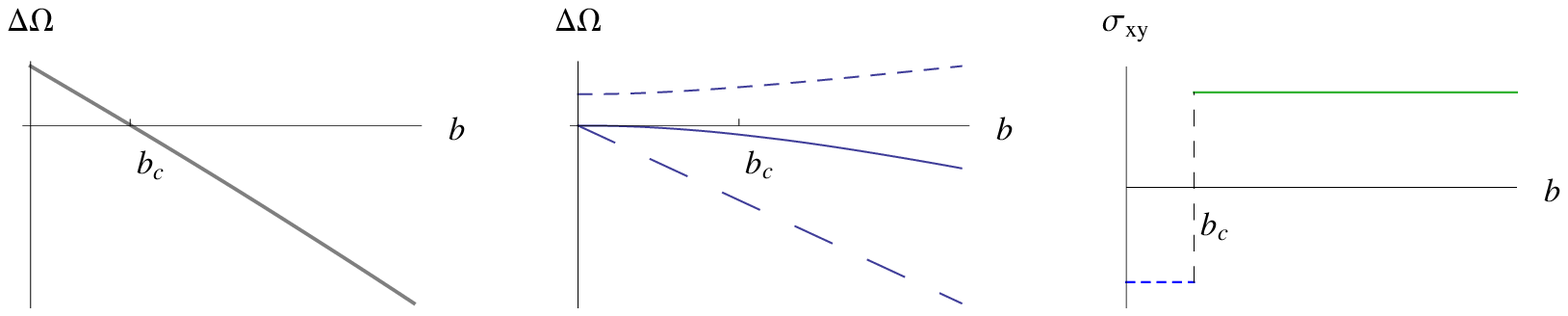}}
%


\newsec{Discussion}

In this paper we propose a holographic model for the quantum Hall plateau transition based on the non--supersymmetric D3/D7 system in the probe limit and at zero temperature.
The full action is obtained by considering the tachyon--DBI action and a Chern--Simons term, which allows the system to have a non--zero charge density.
%
The equations of motion have two  solutions  which are characterized by different values of the action at finite 
chemical potential $\mu$ and magnetic field $b$.
The Chern-Simons term on the world--volume of the flavor D-brane breaks parity;\foot{ This is just a holographic
manifestation of the fact that in a theory of three-dimensional  massless Dirac fermion interacting with the $SU(N)$ gauge field
parity is broken spontaneously \refs{\NiemiRQ,\RedlichKN,\SemenoffDQ}.}
 as a result, there is a first order phase transition
at  $b=0$ .
At this point, the Hall conductivity experiences a jump between the two plateaux.
Of course, in the real quantum Hall setup the physics at small magnetic field is entirely classical.
One may still hope that some universal features of the transition are correctly captured by our model;
we leave detailed investigation of this for future work.

By playing with the holographic action [changing $V(\theta)$ and $c(\theta)$], and breaking parity explicitly,
we can also make phase transitions happening at finite values of the magnetic field.
Note that our description requires introduction of the cutoff,
but can 
be rendered renormalizable by taking the physical limit, where the cutoff $\Lambda$ is taken to infinity and the tachyon mass to the BF 
value, while the physical scale remains fixed.
We expect the physics of the phase transition described in this paper
to not be significantly affected by this procedure.
At finite $\Lambda$ there can be multiple solutions with the same value of the cutoff which oscillate
around $\theta=0$.
We show that their (Gibbs) energy is always higher than those of the solutions we discuss in the paper and hence they
are suppressed thermodynamically.
All these solutions disappear in the physical limit.

We also considered another possibility to generate phase transitions phenomenologically.
Suppose the tachyon potential  has two minima, $\theta_1$ and $\theta_2$, and both satisfy $V(\theta_{1,2})=0$ condition, to
assure the absence of external forces (finite energy density) at $r=r_0$.
Can we have two distinct solutions which interpolate between $\theta=0$ at $r=\Lambda$ and $\theta=\theta_{1,2}$ at $r=r_0$?
Then, there can be a competition between their energies, and, possibly, a phase transition.
We show that at least within the set of examples we considered  two solutions do not arise -- the tachyon field always stops at the
first minimum.

One of the motivations for using the language of a non-linear tachyon action to describe QHE has been its ability to model both the hard-wall and the soft-wall behaviour.
The latter opens up a possibility of describing a crossover between different Hall plateaux at finite temperature 
(the class of the hard-wall models, to which all currently available holographic quantum Hall models belong, is not suitable for this purpose: we expect temperatures much lower than the hard wall scale to not affect the order of the phase transition).
We leave investigation of the holographic quantum Hall in a soft-wall type model for future work.
%

%


\newsec{Acknowledgments}

We thank
A.~Amoretti,
D.~Arean,
R.~Argurio,
O.~Bergman,
A.~Braggio,
R.~Davison,
P.~Fonda,
V.~Giangreco Puletti,
V.~Giraldo-Rivera,
M.~Jarvinen,
N.~Jokela,
S.~Klug,
C.~Kristjansen,
M.~Kulaxizi,
A.~Marzolla,
D.~Musso,
G.~Policastro and
J.~Slingerland
for useful discussions.
We also thank O.~Bergman and V.~Filev for comments on the manuscript.
This work was supported in part by the vidi grant from the NWO.


\listrefs
\bye